\documentstyle[proceedings,epsf]{crckapb}

\def\lesssim{\mathrel{\hbox{\rlap{\hbox{\lower4pt\hbox{$\sim$}}}\hbox{$<$}}}}
\def\plotfiddle#1#2#3#4#5#6#7{\centering \leavevmode
\vbox to#2{\rule{0pt}{#2}}
\includegraphics{#1}}

\begin{opening}
\title{Aluminum Abundances on the Red Giant Branch: Testing Helium Mixing as
 a Second Parameter}
\author{R. M. Cavallo}
\institute{NASA/Goddard Space Flight Center, Greenbelt, MD, USA}
\end{opening}

\begin{document}

Although globular cluster horizontal-branch (HB) morphology is determined
 primarily by the cluster metallicity, observations have shown that clusters
 with similar metallicities sometimes have very different HB's, indicating the
 existence of a second parameter.
Sweigart (1997, 1999) suggested that deep mixing on the red giant branch (RGB)
 may be a second parameter if He is mixed into the stellar envelope.
A He-mixed RGB star would evolve onto the HB at a bluer color and brighter
 luminosity than an unmixed star.
Cavallo {\it et al.} (1998) showed that Al is produced in large quantities only
 inside the H-burning shell where He is synthesized, so that Al makes a good 
 tracer of He.
Thus, a relationship should exist between the ratio of
 Al-rich to Al-normal stars on the RGB and the ratio of blue to red HB stars.
Since He cannot be directly measured in cool giants, Al enhancements can 
 be used as an indicator of He mixing.

In Figures 1a and 1b, we present the Al abundance data for the blue-HB cluster
 M13 and the red-HB cluster M3, both of which have [Fe/H] ${\sim} -1.5$.
The distributions of the M13 and M3 giants are both
 bimodal with a ratio of 2:1 (Al strong to Al poor).
The Al abundances in three M13 giants and all six M3 giants are from Cavallo 
 \& Nagar (1999, CN99) and were derived using the models of Kraft {\em et al.}
 (1992, 1993, 1995). 
The the rest of the M13 data are from the Kraft papers and from Pilachowski
 {\em et al.} (1996).
The similarity between the distributions is surprising given the difference 
 between the two HB's, although the M3 sample is very small and its Al-rich
 stars are not as enhanced as the M13 giants.

\begin{figure}
\plotfiddle{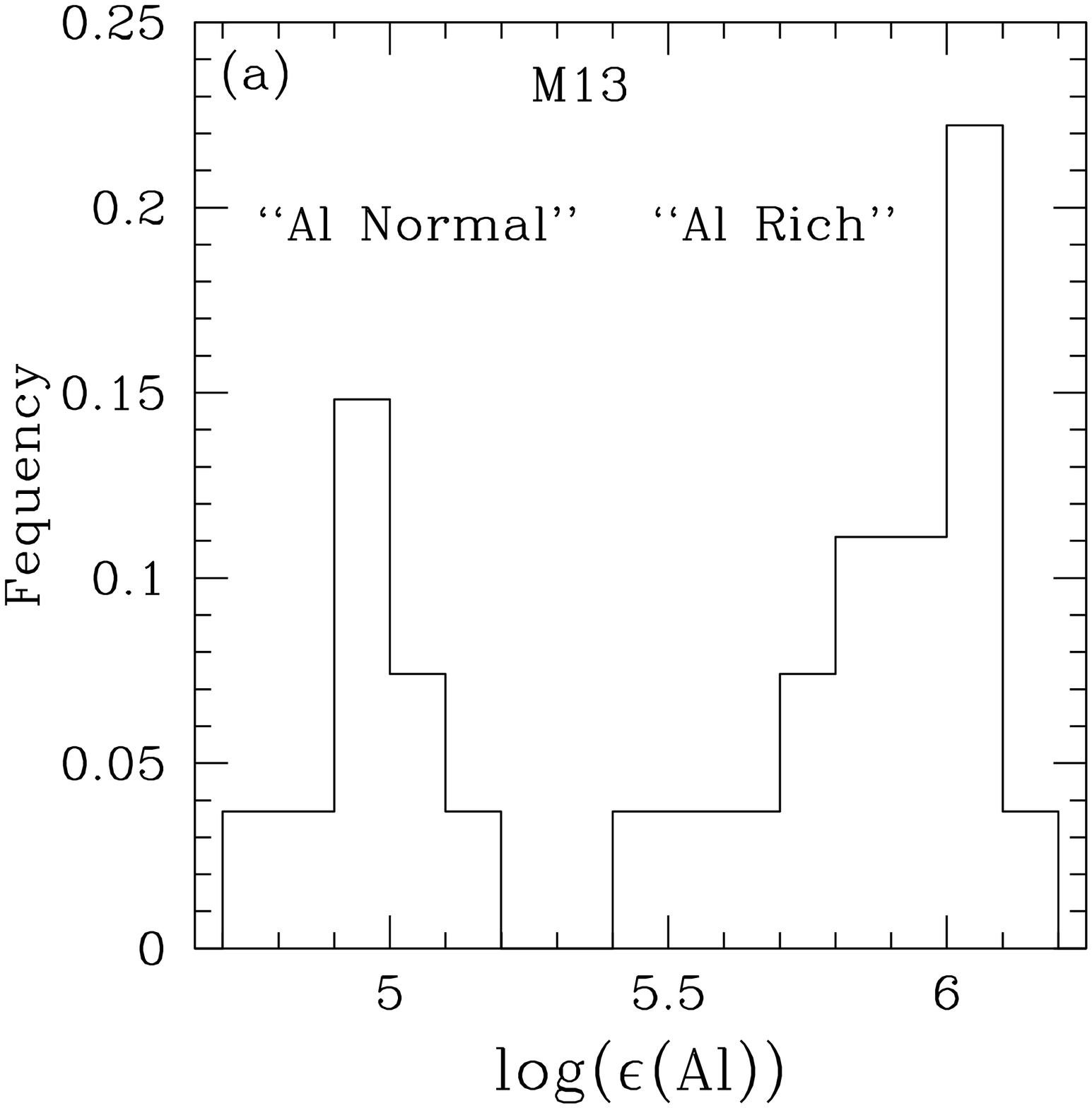}{70pt}{0}{30}{30}{-175}{-125}
\plotfiddle{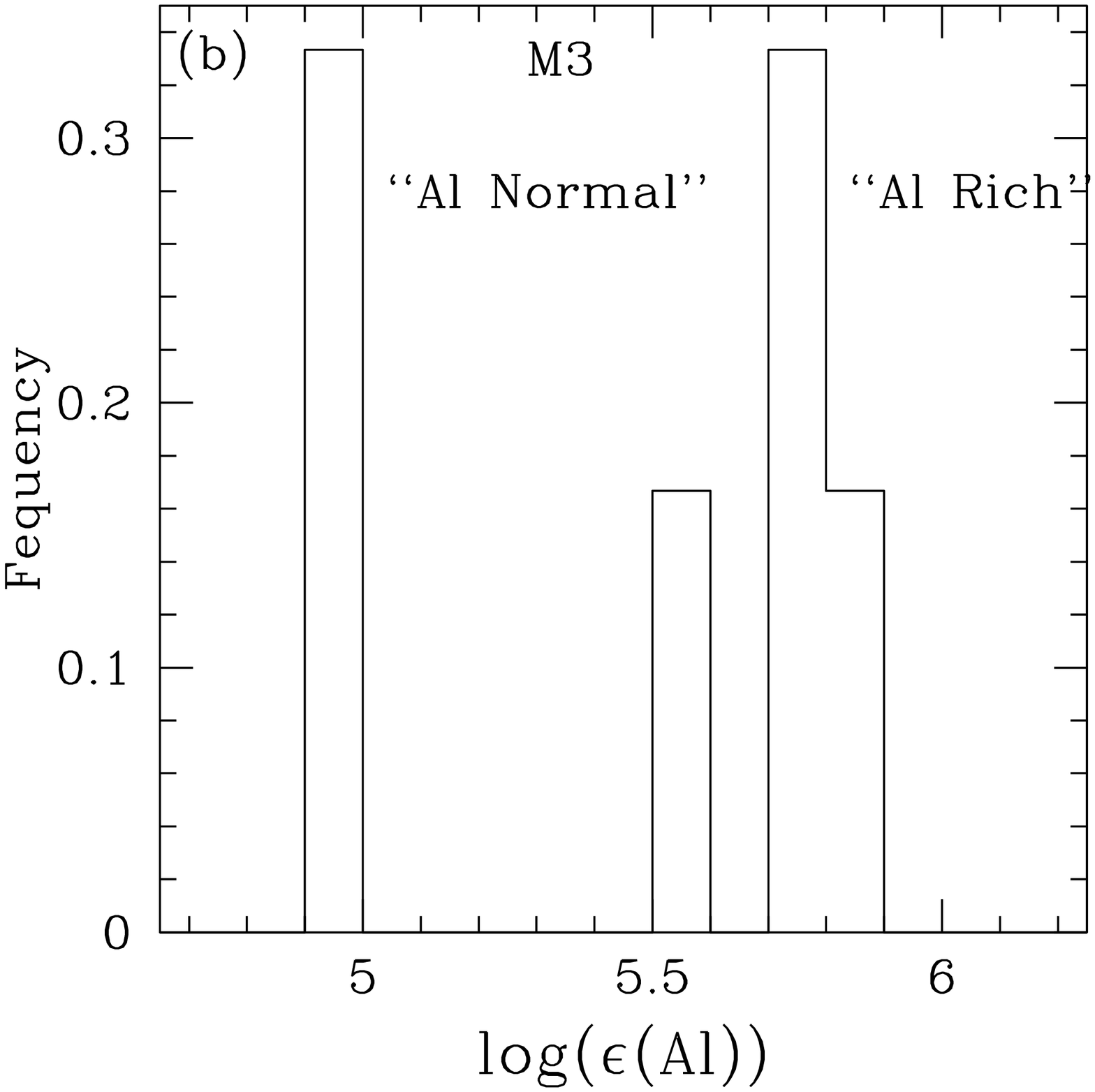}{70pt}{0}{30}{30}{0}{-42}
\caption{Al abundance distribution for (a) 27 M13 and (b) 6 M3 giants.}
\end{figure}

Ferraro {\em et al.} (1998) used HST data to show that the empirical ratio
 of blue to red HB stars is 58:42 in M13; i.e., close to the ratio of
 Al-strong to Al-normal giants.
No similar HST data exist for M3, although it does have a redder HB morphology.
While the data might support the He-mixing hypothesis for M13,
 they suffer from two shortcomings:
 (1) the sample sizes of RGB stars with Al measurements is
 small, even for M13, and (2) some giants are likely to have been
 contaminated by an early generation of massive asymptotic-giant-branch
 (AGB) stars that evolved quickly before shedding their processed envelopes
 into the cluster.
If the AGB stars are greater than ${\sim} 4 \ {\rm M}_{\odot}$, it is likely
 that hot-bottom-burning products, such as Al, would be ejected along with 
 $s$-process elements, such as Ba (Lattanzio {\em et al.} 1997).
Thus, looking for a relationship between Ba and Al, for example, might help
 to discriminate between stars that have been contaminated and those that
 are truly self-enriched; i.e., giants with both enhanced Al and Ba
 could be removed from the sample, making the connection between He mixing
 and Al clearer.
That such a relationship might exist is suggested by the data of Ivans 
 {\em et al.} (1999, this issue) who show that for 20 giants in M4
 ([Fe/H] ${\sim} -1$) the Al and Ba abundances are both elevated on average.
Unfortunately, they do not report on individual stars, so it is difficult to
 determine if a correlation exists or not.
Such a relationship might, however, explain the M3 giants with Al enhancements.

\end{document}